\documentclass[aps,prb,twocolumn,superscriptaddress]{revtex4}

\usepackage{graphicx}
\usepackage{bm}
\newcommand{\unit}[1]{\;\mathrm{#1}}

\begin{document}
\title{High Domain Wall Velocity at Zero Magnetic Field Induced by Low Current Densities in Spin Valve Nanostripes}

\author{Stefania Pizzini}
\affiliation{Institut N\'{e}el, Centre National de la Recherche
Scientifique (CNRS) and Universit\'{e} Joseph Fourier, B.P.166,
38042 Grenoble, France}
\author{Vojt\v ech Uhl\'\i\v r}
\affiliation{Institut N\'{e}el, Centre National de la Recherche
Scientifique (CNRS) and Universit\'{e} Joseph Fourier, B.P.166,
38042 Grenoble, France} \affiliation{Institute of Physical
Engineering, Brno University of Technology, 61669 Brno, Czech
Republic}
\author{Jan Vogel}
\affiliation{Institut N\'{e}el, Centre National de la Recherche
Scientifique (CNRS) and Universit\'{e} Joseph Fourier, B.P.166,
38042 Grenoble, France}
\author{Nicolas Rougemaille}
\affiliation{Institut N\'{e}el, Centre National de la Recherche
Scientifique (CNRS) and Universit\'{e} Joseph Fourier, B.P.166,
38042 Grenoble, France}
\author{Sana Laribi}
\affiliation{Unit\'{e} Mixte de Physique CNRS/Thales and
Universit\'{e} Paris Sud 11, 91767 Palaiseau, France}
\author{Vincent Cros}
\affiliation{Unit\'{e} Mixte de Physique CNRS/Thales and
Universit\'{e} Paris Sud 11, 91767 Palaiseau, France}
\author{Erika Jim\'{e}nez}
\affiliation{Dpto. F\'{i}sica de la Materia Condensada and Inst.
Nicol\'{a}s Cabrera, Universidad Aut\'{o}noma de Madrid, 28049
Madrid, Spain}
\author{Julio Camarero}
\affiliation{Dpto. F\'{i}sica de la Materia Condensada and Inst.
Nicol\'{a}s Cabrera, Universidad Aut\'{o}noma de Madrid, 28049
Madrid, Spain} \affiliation{Instituto Madrile\~{n}o de Estudios
Avanzados en Nanociencia (IMDEA-Nano), Campus de Cantoblanco, 28049
Madrid, Spain}
\author{Carsten Tieg}
\affiliation{European Synchrotron Radiation Facility, rue Jules
Horowitz, BP200, 38043 Grenoble, France}
\author{Edgar Bonet}
\affiliation{Institut N\'{e}el, Centre National de la Recherche
Scientifique (CNRS) and Universit\'{e} Joseph Fourier, B.P.166,
38042 Grenoble, France}
\author{Marlio Bonfim}
\affiliation{Departamento de Engenharia El\'{e}trica, Universidade
Federal do Paran\'{a}, 81531-990 Curitiba PR, Brasil}
\affiliation{Institut N\'{e}el, Centre National de la Recherche
Scientifique (CNRS) and Universit\'{e} Joseph Fourier, B.P.166,
38042 Grenoble, France}
\author{Richard Mattana}
\affiliation{Unit\'{e} Mixte de Physique CNRS/Thales and
Universit\'{e} Paris Sud 11, 91767 Palaiseau, France}
\author{Cyrile Deranlot}
\affiliation{Unit\'{e} Mixte de Physique CNRS/Thales and
Universit\'{e} Paris Sud 11, 91767 Palaiseau, France}
\author{Fr\'{e}d\'{e}ric Petroff}
\affiliation{Unit\'{e} Mixte de Physique CNRS/Thales and
Universit\'{e} Paris Sud 11, 91767 Palaiseau, France}
\author{Christian Ulysse}
\affiliation{CNRS, PhyNano Team, Laboratoire de Photonique et de
Nanostructures, route de Nozay, F-91460 Marcoussis, France}
\author{Giancarlo Faini}
\affiliation{CNRS, PhyNano Team, Laboratoire de Photonique et de
Nanostructures, route de Nozay, F-91460 Marcoussis, France}
\author{Albert Fert}
\affiliation{Unit\'{e} Mixte de Physique CNRS/Thales and
Universit\'{e} Paris Sud 11, 91767 Palaiseau, France}


\begin{abstract}
Current-induced magnetic domain wall motion at zero magnetic field
is observed in the permalloy layer of a spin-valve-based nanostripe
using photoemission electron microscopy. The domain wall movement is
hampered by pinning sites, but in between them high domain wall
velocities (exceeding 150 m/s) are obtained for current densities
well below $10^{12} \unit{A/m^2}$, suggesting that these trilayer
systems are promising for applications in domain wall devices in
case of well controlled pinning positions. Vertical spin currents in
these structures provide a potential explanation for the increase in
domain wall velocity at low current densities. \end{abstract}

\maketitle

Moving magnetic domain walls using electric currents via spin-torque
effects rather than using a magnetic field is one of the recent
exciting developments in spintronics~\cite{Marrows2005}. Since the
prediction of spin-torque effects~\cite{Berger1984}, many
experimental~\cite{Grollier2003,
Yamaguchi2004,Lim2004,Klaui2005,Ravelosona2005,Klaui2006,Yamanouchi2007,Laribi2007,Hayashi2007a,Tanigawa2008,Vanhaverbeke2008,Koyama2008,Moore2008}
and theoretical~\cite{Tatara2004,Li2004b,Thiaville2005} works have
been dedicated to the study of current-induced domain wall motion
(CIDM). Besides fundamental investigations, the use of domain walls
in logic~\cite{Allwood2005} and memory~\cite{Parkin2008} devices has
already been proposed. Low current densities and high domain wall
(DW) velocities at zero magnetic field are required for future
applications.

Direct evidence of CIDM at zero field has been reported for several
nanostripe systems, including permalloy
(FeNi)~\cite{Klaui2005,Hayashi2007a}, magnetic
semiconductors~\cite{Yamanouchi2007} and systems with perpendicular
magnetization~\cite{Ravelosona2005,Tanigawa2008,Koyama2008,Moore2008}.
For the commonly used FeNi system, the critical current densities
are not much below $10^{12} \unit{A/m^2}$ at zero magnetic
field~\cite{Klaui2006,Hayashi2007a}, associated with DW velocities
going from some m/s up to about $100
\unit{m/s}$~\cite{Hayashi2007a}. Much lower critical currents are
found for magnetic semiconductors like GaMnAs (about $1\times10^9
\unit{A/m^2}$) because of the low magnetic moments, but the observed
DW velocities are small ($< 1 \unit{m/s}$)~\cite{Yamanouchi2007}.
Moreover, these materials are not ferromagnetic at room temperature.
Low current density values are also found in spin-valve-based
nanostripes with either in-plane~\cite{Grollier2003,Laribi2007} or
perpendicular anisotropy~\cite{Ravelosona2005}. Additionally,
transport measurements in FeNi/Cu/Co trilayers show CIDM induced by
subnanosecond current pulses~\cite{Lim2004}, indirectly indicating
high DW velocities in such spin-valve-based systems. In this work,
we show that in these systems CIDM at zero magnetic field can take
place with high DW velocities (exceeding 150 m/s) at current
densities well below $10^{12} \unit{A/m^2}$. These high velocities
are observed only in  certain regions of the nanostripes, where
domain wall pinning is limited. Currents perpendicular to the plane
in the vicinity of the DW are probably partly responsible for this
increase in efficiency, which makes the trilayer systems possible
candidates for spintronic applications based on CIDM if pinning can
be controlled.

\begin{figure}[ht!]
\includegraphics*[bb= 284 437 415 545]{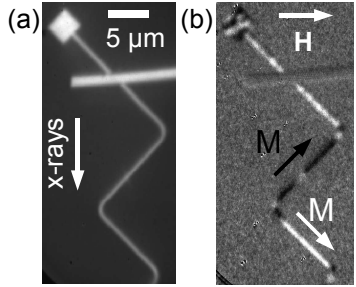}
\caption{Topographic \textbf{(a)} and magnetic \textbf{(b)} PEEM
images of a 400~nm wide spin-valve-like nanostripe with a zigzag
angle of 90$^\circ$, taken at the Ni L$_3$-edge. A static in-plane
magnetic field of 400~mT was applied perpendicular to the incoming
X-ray beam before the measurements, as indicated by the arrow at the
top of \textbf{(b)}. In \textbf{(b)}, white (black) contrast
corresponds to domains having a magnetization component parallel
(antiparallel) to the incoming x-ray direction, as indicated by the
black and white arrows. The horizontal bar in the upper part of the
images is one of the gold contacts for current injection.}
\label{fig:DWcorners}
\end{figure}

We observed domain wall motion in the FeNi layer of 400~nm wide
FeNi~(5~nm)/Cu~(8~nm)/Co~(7~nm)/ CoO~(3~nm) nanostripes, by using
Photoemission Electron Microscopy (PEEM) combined with X-ray
Magnetic Circular Dichroism (XMCD) \cite{Locatelli2008}. The Cu
spacer layer is chosen to be sufficiently thick to avoid any
interlayer exchange coupling or magnetostatic orange-peel coupling
between the magnetic layers. The nanostripes were processed from
film stacks grown by sputtering on high resistivity Si substrates,
combining electron beam lithography and lift-off technique. They
were patterned in zigzag shapes with angles of 90 and 120$^{\circ}$
[see Fig.~\ref{fig:DWcorners}(a)]. Contact electrodes made of Ti/Au
were subsequently deposited using evaporation and lift-off
technique.

The samples were mounted on a sample holder that allows both
magnetic field pulses and current pulses to be applied. For field
pulses, we used a combination of double stripline-like microcoils
and a home-made pulsed current supply~\cite{Vogel2004}. Nanosecond
magnetic field pulses could be applied in the plane of the stripes,
perpendicular to the long axis of the zigzag (see
Fig~\ref{fig:DWcorners}). Current pulses were injected into the
nanostripes using a generator providing voltage pulses with 4 ns
risetime and tunable length. The current flowing through the stripes
was deduced from the voltage measured over the 50~$\Omega$ entrance
of a 6 GHz oscilloscope connected in series with the stripes.

Our XMCD-PEEM measurements were performed at the European
Synchrotron Radiation Facility beamline ID08. In order to avoid
discharges from the objective lens of our Focus IS-PEEM, which is at
a distance of $2 \unit{mm}$ from the sample, the voltage on the
objective lens was kept below 4.2 keV, limiting the spatial
resolution to about 0.7 $\mu$m. In order to image the domain
structure in the FeNi layer, the x-ray energy was tuned to the Ni
$L_3$ absorption edge (852.8 eV). To optimize the magnetic contrast,
the difference between two consecutive images obtained with $100\%$
left- and right-circularly polarized x-rays was computed. The
presence of a rather thick Cu spacer ($8 \unit{nm}$) layer, combined
with the limited escape depth of the secondary electrons, prevented
images of the Co domain structure to be taken.

Before introducing the samples into the PEEM microscope, we applied
an in-plane magnetic field of 400 mT perpendicularly to the zigzag
long axis. The remanent magnetic state obtained for a 400~nm wide,
90$^\circ$ angle zigzag nanostripe is shown in
Fig.~\ref{fig:DWcorners}(b). The contrast is given by the projection
of the magnetization on the beam direction, i.e., white (black)
domains have their magnetization pointing downwards (upwards), along
the stripes. The magnetization is nearly saturated in the straight
sections, but a black-white-black-white contrast is visible at the
bends. The magnetization in the straight sections of the buried Co
layer is expected to be parallel to the FeNi magnetization, leading
to domain walls at the bends. These head-to-head or tail-to-tail
domain walls lead to strong magnetostatic effects that induce a
local antiparallel alignment between Co and FeNi layers in the
vicinity of the Co DW \cite{Vogel2007}. In the straight sections,
these magnetostatic interactions caused by Co domain walls should
not be present.

\begin{figure}[ht!]
\begin{center}
\includegraphics*[bb= 159 370 364 534]{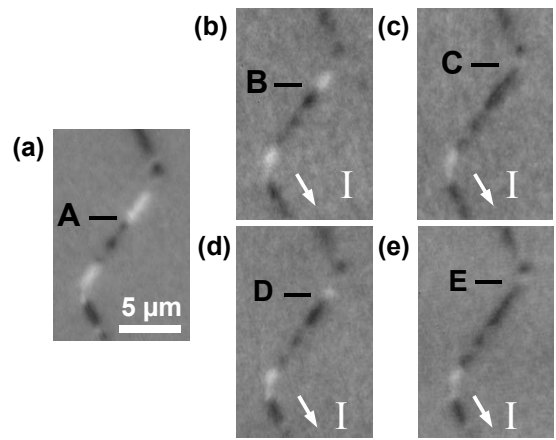}
\end{center}
\caption{XMCD-PEEM images of the FeNi layer of a 400~nm wide
spin-valve-like nanostripe with a zigzag angle of 120$^\circ$.
\textbf{(a)} gives the initial domain state, obtained after applying
an in-plane 50 mT magnetic field pulse perpendicular to the long
direction of the stripe. \textbf{(b)} and \textbf{(c)} show the
images after application of one, resp. two 100 ns long current
pulses of $+ 2 \unit{mA}$ starting from the domain state of
\textbf{(a)}. The indicated DW moves from position A to position B
with the first pulse, and from B to C with the second pulse.
\textbf{(d)} and \textbf{(e)} show images obtained after application
of one 15 ns pulse of $+ 5 \unit{mA}$ and one 20 ns pulse of $+ 5
\unit{mA}$, respectively, in both cases starting from the initial
state of \textbf{(a)}. The 15 ns pulse makes the DW move from
position A to D, the 20 ns pulse causes a motion from A to E.}
\label{fig:5.4V}
\end{figure}

In Fig.~\ref{fig:5.4V}, we show current-induced DW motion in a
400~nm wide stripe with zigzag angles of 120$^\circ$. The initial
domain structure was obtained after application of several magnetic
field and current pulses. Domain wall pinning in these structures is
quite strong, and several attempts to induce domain wall motion with
current pulses were made before reproducible motion could be
obtained. The domain structure in Fig.~\ref{fig:5.4V}(a) could be
obtained reproducibly by applying 50 ns long magnetic field pulses
in the plane of the samples, to the left in the figure, with an
amplitude of 50 mT. Starting from this initial state, we applied
current pulses with different amplitudes and lengths, in order to
determine the DW velocity and the threshold current in this section.
Figure~\ref{fig:5.4V}(b) shows the domain structure obtained after
applying one 100 ns long current pulse with an amplitude of $+ 2
\unit{mA}$, a value below which no DW motion was detected for these
relatively short pulses. This threshold current pulse causes a
displacement of the domain wall from position A to position B in the
images. A consecutive pulse with the same amplitude and length
induces a further movement of the same domain wall in the same
direction, from B to C [Fig.~\ref{fig:5.4V}(c)]. Note that in these
images only one DW moves for the applied amplitude of the current,
showing that the pinning strengths can strongly differ at different
positions in the nanostripe. The free motion of the FeNi DW over
section A-C in Fig.~\ref{fig:5.4V} indicates that pinning in this
section is relatively small.

The first pulse causes a CIDM of $(1.75 \pm 0.2) \unit{\mu m}$, the
second pulse $(1.92 \pm 0.2) \unit{\mu m}$, resulting in a domain
wall velocity of about $(18 \pm 2)\unit{m/s}$. The current density
in the FeNi layer corresponding to the $+ 2 \unit{mA}$ pulse is $2
\times 10^{11} \unit{A/m^2}$ if we consider a uniform current
distribution through the trilayer stack and $4 \times 10^{10}
\unit{A/m^2}$ if we suppose that the current density is proportional
to the conductivity in each layer~\cite{Laribi2007}. The value of $2
\times 10^{11} \unit{A/m^2}$ gives thus an upper bound for the
current density in the FeNi layer. In Figs.~\ref{fig:5.4V}(d) and
\ref{fig:5.4V}(e), we show results of measurements using current
pulses with the maximum available amplitude of $5 \unit{mA}$
(corresponding to a current density of $5 \times 10^{11}
\unit{A/m^2}$ in the FeNi layer, or $1 \times 10^{11} \unit{A/m^2}$
for non-uniform current distribution) and lengths of 15 and 20 ns,
in both cases starting from the initial state shown in
Fig.~\ref{fig:5.4V}(a). The CIDM for the 15 and 20 ns pulses are
$(2.7 \pm 0.2) \unit{\mu m}$ and $(3.4 \pm 0.2) \unit{\mu m}$,
showing that in this section of the nanostripe the displacement is
about proportional to the pulse length. The resulting DW velocities
are $(180 \pm 10)$ and $(170 \pm 10)\unit{m/s}$, respectively. This
result clearly shows that for this trilayer system the
current-induced domain wall velocities between pinning sites are
above literature values for single FeNi layers~\cite{Hayashi2007a},
for current densities that are at least a factor two smaller. The
above results for strongly different pulse lengths indicate that for
this section the displacements are proportional to the pulse length.
This suggests that DW movement takes place only during the current
pulse and that movements after the current pulse, for instance due
to attractive potential wells, are small. Domain wall pinning can be
caused by edge roughness induced by the lithography, stray fields of
domain walls in the Co layer, interface roughness or grain
boundaries. Scanning electron microscopy of our structures indicates
a lateral roughness smaller than 1 nm. Roughness of the Cu/FeNi
interface should be small with respect to the total FeNi layer
thickness of 5 nm. Pinning at grain boundaries could be reduced
using amorphous layers of CoFeB instead of FeNi~\cite{Laribi2007}.

In general, experimental results of current-induced DW dynamics are
described (at least qualitatively) using a phenomenological model
with two components for the spin transfer torque, i.e., the
adiabatic and the non-adiabatic torque. In the simplest approach of
a 1D domain wall model~\cite{Thiaville2005}, the DW velocity is
expressed as $v_{DW}=\frac{\beta}{\alpha}u$ where $u$ is the rate of
spin angular momentum transfer from the conduction electrons to the
local moments in the domain wall [$u = JPg\mu_B/(2eM_S)$ where $J$
is the current density and $P$ its polarization
rate~\cite{Thiaville2005}]. Several experiments indicate a value of
the non-adiabaticity parameter $\beta$ close to $\alpha$, the
damping term~\cite{Beach2006,Meier2007}. However, recently Hayashi
\textit{et al.}~\cite{Hayashi2007a} observed a maximum DW velocity
of 110 m/s (for $J = 1.5\times10^{12} \unit{A/m^2}$) at zero
magnetic field in 300 nm FeNi nanostripes. This velocity is larger
than the spin angular momentum transfer rate $u$ for any $P < 1$
(for $P=0.7$, $u$ is about 75 m/s and $\frac{\beta}{\alpha} \approx
1.5$). Using the same parameters, our current density $J \approx
5\times10^{11} \unit{A/m^2}$ would lead to a spin transfer rate $u$
of about 25 m/s. The only way to explain our measured velocity of
175 m/s within this model would be to assume a
$\frac{\beta}{\alpha}$ of about 7. Such a very large
non-adiabaticity seems unlikely and other spin transfer mechanisms
must be considered to explain our high DW velocities.

A specificity of our spin-valve structures is that part of the
incident spin flux in the FeNi layer is transformed, in the region
around the DW, into spin accumulation in the Cu and Co layers. This
spin accumulation in the Cu spacer layer below the DW induces a
vertical (Current Perpendicular to the Plane or CPP) spin current
that flows into the DW and exerts an additional torque on the DW.
This situation is somewhat similar to that of spin transfer in the
CPP geometry of magnetic pillars. Our results suggest that the yield
of this additional vertical spin transfer channel is much higher
than the yield of the in-plane spin transfer channel, as predicted
recently by micromagnetic simulations on similar nanostructures
containing a single DW~\cite{Khvalkovskiy2008}. Our direct
observation of high velocity current-induced DW motion confirms the
potential of these trilayer systems for applications in DW-based
magnetic memories and logic devices.

\acknowledgements We acknowledge the technical support of E.~Wagner,
P.~Perrier, D.~Lepoittevin, and L.~Delrey, as well as the
experimental help of S.~Pairis, T.~Fournier, and W.~Wernsdorfer. We
are grateful to A.~Khvalkovskiy, A.~Anane, and J.~Grollier for
discussions. E.J. and J.C. acknowledge financial support through
projects HF2007-0071, MAT2006-13470, and CSD 2007-00010. V.U. was
financially supported by the grants No. MSM0021630508 and No.
KAN400100701. This work was partially supported by the
ANR-07-NANO-034 `Dynawall'.

\end{document}